%Paper: hep-ph/9209260
%From: U6405580@ucsvc.ucs.unimelb.edu.au
%Date: 21 Sep 1992 15:52:27 +1000

\magnification=1200
{\nopagenumbers
\baselineskip=20pt
\hsize=5.5in
\hskip10cm UM--P-92/75\hfil\par
\hskip10cm OZ-92/26\hfil
\vskip2cm
\centerline{ CP Violating Form Factors for Three Gauge Boson Vertex} \par
\centerline{in the Two Higgs Doublet and Left-Right symmetric Models}\par
\vskip2cm
\hskip4cm X. G. He, J. P. Ma and B. H. J. McKellar\par
\vskip0.5cm
\hskip4cm Recearch Center for High Energy Physics \par
\hskip4cm School of Physics\par
\hskip4cm University of Melbourne \par
\hskip4cm Parkville, Victoria 3052\par
\hskip4cm Australia \par
\vskip2cm
{\bf\underbar{Abstract}}:
In this paper we calculate the one loop contributions to the CP violating
three gauge boson couplings in two-Higgs doublet and Left--Right symmetric
models. In the two-Higgs doublet model only a P conserving and CP violating
coupling is generated, and it can be large as $10^{-3}$. In the Left--Right
symmetric model both P conserving and violating couplings are generated.
Due to constraints on the $W_L$--$W_R$ mixing, these couplings are small.
\par\vfil\eject }
\baselineskip=15pt
\pageno=1
At $e^+ e^-$ colliders with enough energy $W$ pairs can be produced.
Such facilities can be used to detect the three gauge boson couplings
and  provide tests for the Standard Model (SM). An important feature of $e^+
e^-$ colliders is that they give very nice opportunities to look for CP
violation because the initial state is CP eigenstate.
There have been many studies on the anomalous CP conserving couplings
 of three gauge bosons$[1,2]$. In this paper we will concentrate on
possible CP violating $VW^+W^-$
(here $V$ stands for $Z$ or $\gamma$) three gauge
boson couplings in some models. Under $U(1)_{em}$-invariance the CP violating
vertex for $VWW$ can be written with form factors(we use the same convention
 as in $[1]$):
$$ \eqalign{
\Gamma_{\mu_1\mu_2 \nu} =& if_4^V(q_{\mu_2}g_{\mu_1\nu} + q_{\mu_1}g_{\mu_2
 \nu}) \cr
+& f_6^V\varepsilon_{\mu_1\mu_2\nu\rho}q^{\rho} + {1\over m_W^2}f_7^V
\varepsilon_{\mu_1\mu_2\sigma\rho}(k_1-k_2)_{\nu}(k_1-k_2)^{\sigma}q^\rho
 \cr} \eqno(1) $$
where $q$, $k_1$ and $k_2$ are the momenta of $V$, $W^+$ and $W^-$ bosons
respectively. The $W$ bosons are on shell.
The $f_i^V$ are the form factors depending on  $q^2$ in general.
Of these, $f_4^V$ is P conserving and CP
violating whereas $f_6^V$ and $f_7^V$ are P and CP violating.\par
The form factors $f_6^\gamma$ and $f_7^\gamma$ can be constrainted by using low
energy
data[3,4,5]. The best constraints on these form factors at $q^2=0$ are obtained
by relating them to electric dipole moments (EDM) of
electron
and neutron. From the experimental upper bound on the electron EDM, one obtains
[4]: $f_6^\gamma \le 4.1\times 10^{-2}$, $f_7^\gamma \le 1.1\times 10^{-2}$.
The
constraints from the
experimental bound on neutron EDM are less certain, the bound
varies from $10^{-4}$ to O(1) depending on the method used.
Constraints on $f_{6,7}^Z$ have also been obtained[4].
So far,  there is no constraint on $f_4^Z$ from CP odd effects in experiments,
CP even effect one gets $f_4^Z\le 0.24$[5].
Needless to say much effort should be
spent
to measure these important form factors to find CP violation
somewhere other than $K_L^0$ meson system. \par
 In the minimal SM,
 CP
violating three gauge boson couplings are zero up to the two loop level at
least and
therefore are too small to be seen. If CP violation were to be detected
experimenally in the $VWW$ vertex, this would imply new CP
violating interactions beyond the minimal SM.
 We will study the CP violating three gauge boson couplings in two
extensions of the SM, the two-Higgs doublet and
the Left-Right symmetric models.
The form factors $f_{6,7}^\gamma$ at $q^2=0$, which are
related to the  electric dipole moment of $W$ boson,
have been considered in
multi-Higgs doublet models and Left-Right symmetric models[7,8,9].
We will calculate all the form factors in eq. (1) with arbitrary $q^2$.
\par
We consider the two Higgs doublet model first. In this model the gauge group is
the same as the minimal SM , i.e., $SU(3)_C\times SU(2)_L\times U(1)_Y$ and
has two Higgs representations transforming under $SU(2)_L$ as doublet.
It is possible to have CP violation in the Higgs
sector in this model. If such CP violation is allowed, in general there will be
flavour
 changing neutral current (FCNC) at the tree level. There are very strong
experimental constraints on FCNCs. These FCNCs can be avoided by imposing
 certain symmetries such that only one Higgs doublet generates masses for one
class of
given charge fermions. However, if these symmetries are exact for the complete
theory, CP violation in the Higgs sector will also be eliminated. We will take
the approach of  allowing
 sorft symmetry breaking terms in the Higgs potential[10].
Thus we have a theory consistent with experiment. Let $\phi_i^T = (\phi_i^+,
\phi_i^0+v_ie^{i\varphi_i})$ with $i = 1\;,2$ be the two Higgs doublets. Here
$v_ie^{i\varphi_i}$ are their vacuum expectation values. The CP violating
phases
can be chosen such that $\varphi_1 = 0$. In this model, there are three
neutral and one charged physical Higgs particles. It is convenient to work in a
 basis where the would-be Goldstone bosons are decoupled from physical ones. To
this end, we first rotate the Higgs field in a new basis, $\phi_1'^T=(G^+,
H_1^0+iG^0+\sqrt{v_1^2+v_2^2})$ and $\phi_2'^T=(H^+,H^0_2+iI_2^0)$ such that
only the VEV of $\phi_1'$ is not zero. In this basis, the
would-be goldstone bosons are $G^+$ and $G^0$. We have
  $$ \left( \matrix{ H_1^0+iG^0 \cr
                     H_2^0+iI_2^0 \cr} \right) =U \left( \matrix{
        \phi_1^0 \cr \phi_2^0\cr }\right), \ \
       \left( \matrix{ G^+\cr H^+ \cr}\right)=U \left(
    \matrix{\phi_1^+\cr\phi_2^+ \cr} \right) \eqno(2) $$
$U$ is a $2\times 2$ matrix:
   $$ U=\left( \matrix{ \cos\beta & \sin\beta e^{-i\varphi_2} \cr
                        -\sin\beta e^{i\varphi_2} & \cos\beta \cr} \right)
   \eqno(3)$$
where $\cos\beta = {v_1\over \sqrt{v_1^2+v_2^2}}$ and $\sin\beta = {v_2\over
\sqrt{v_1^2+v_2^2}}$. $H^+$ is the mass eigenstate of the physical charged
Higgs particle. In general $H^0_i$ and $I^0_2$ are not mass eigenstates. There
are mixings, and if CP is violated there are mixings between $I^0_2$ and
$H^0_i$. We will parametrize the mass eigenstates $\varphi_{mi}$
in terms of $G^0_2$ and $H^0_i$
by a orthogonal matrix
$d_{ij}$, we have
   $$ \left( \matrix{
H_1^0\cr  H_2^0\cr I_2^0\cr } \right) =
   \left( \matrix{
d_{11}&d_{12}&d_{13}\cr
d_{21}&d_{22}&d_{23}\cr
d_{31}&d_{32}&d_{33}\cr} \right)
 \left(\matrix{
\phi_{m1}\cr \phi_{m2}\cr \phi_{m3}\cr }\right) \eqno(4) $$
At the one loop level, we find that only P conserving and CP violating
$ZW^+W^-$
coupling will be generated. There are two diagrams
 contribute which are shown in Fig. 1.
Evaluating these diagrams we obtain
 $$ \eqalign {
f_4^Z(q^2) =& {e^2\over 64\pi^2 sin^2\theta_W cos^2\theta_W}\epsilon_{ijk}
d_{1i}d_{1j}d_{1k} \cr
\cdot & \int [d\alpha](\alpha_1-\alpha_2)\ln{-\alpha_1\alpha_2 q^2 -
 \alpha_3(1-\alpha_3)m_W^2 + \alpha_1 m_j^2
+\alpha_2m_i^2 + \alpha_3m_W^2 \over -\alpha_1\alpha_2q^2-\alpha_3(1-\alpha_3)
m_W^2 + \alpha_1m_j^2 + \alpha_2m_i^2 + \alpha_3 m_C^2} \cr} \eqno(5)$$
where $q$ is the momentum carried by $Z$, $m_W$, $m_i$ and $m_C$ are the
$W$, neutral Higgs and charged Higgs masses
respectively, and $\int [d\alpha] = \int^\infty_0 d\alpha_1 d\alpha_2 d\alpha_3
\delta(1-\alpha_1-\alpha_2-\alpha_3)$.
\par
When $q^2$ is large enough an absorptive part of the form factor will be
generated, which is given by
  $$ \eqalign {
 {\rm Im} f^Z_4
=& {e^2\over 2 \sin^2\theta_W \cos^2\theta_W}\epsilon_{ijk}d_{1i}d_{1j}
d_{1k} \cr
&\times \theta(q^2 - (m_i +m_j)^2)[f(m_i^2,m_j^2, m_C^2)
- f(m_i^2,m_j^2,m_W^2)] \cr
f(x,y,z) =& {y-x\over 256\pi q^6\beta^3_W}\{[\beta_W^4q^4 +
\beta^2_W(8zq^2 - 2q^4
+4q^2(x+y) - 4(x-y)^2)\cr
 +& 16z^2 +8zq^2 -16z(x+y) + q^4 - 4q^2(x+y) + 4(x+y)^2]
\ln {a+b\over a-b} \cr
  +& 8 b(\beta_W^2q^2 + 4 z + q^2 - 2(x+y))\} \cr } \eqno(6)$$
where
 $$ \eqalign {
\beta_W =& \sqrt{1 - {4m_W^2\over q^2}} \cr
 a =& m_W^2 + {1\over 2}(x+y) - z - {1\over 2}q^2 \cr
b=& {1\over 2}\beta_W\sqrt{(q^2 -x+y)^2 -4yq^2} \cr } \eqno(7) $$
It is easy to see from eq.(5) that the CP violating form factor is
proportional to $d_{11}d_{12}d_{13}$ as
indicated in [11]. Further, if the charged Higgs has the same mass as $W$,
$f_4^Z$ is zero. For our numerical results for the form factor we assume
that the Higgs particle $\phi_{m3}$ is too heavy to give sizeable
contribution. The form factor is then a function of $m_1$, $m_2$, $q^2$ and
$m_C$.
In fig.2, we
show some typical values for $f_4^Z$ for different Higgs masses,
where $m_1=60$GeV. The magnitude
also
depends on the CP violating parameter $d_{11}d_{12}d_{13}$. The
maximum possible
value for this quantity is $1/3\sqrt{3}$. So far the best constrain on
these CP violating
parameters in two Higgs doublet model are from fermion dipole moment
constraint.
One of the biggest contributions to the neutron EDM is from the
quark colour dipole
moment. The contribution from up and down quark
color dipole moments is given in the $SU(6)$ approximation[12] by:
  $$ \eqalign {
        D_n  \approx & 2.5\times 10^{-26}{\rm ecm}\{ (d_{1j}-\tan\beta d_{2j}
    )\tan\beta d_{3j}(f({m_t^2 \over m_j^2})+ g({ m_t^2\over m_j^2})) \cr
   & + {m_d \over 2m_u}((d_{1j}-\tan\beta d_{2j}){\rm ctan}\beta d_{3j}
     f({m_t^2\over m_j^2})+(d_{1j}+{\rm ctan}\beta d_{2j})\tan\beta d_{3j}
   g({ m_t^2\over m_j^2}))\} \cr} \eqno (8) $$
which has to be less than the experimental upper bound of $10^{-25}$ecm.[13].
The functions $f(z)$ and $g(z)$ are defined in [12].
 The strange quark color dipole moment gives
a similar constraint[14].
The maximum value for $d_{11}d_{12}d_{13}$ is not ruled out. The P conserving
and CP violating form factor in the two Higgs model can be as large as
$10^{-3}$.
\par
Let us now study the CP violating three gauge boson form factors in Left-Right
symmetric models. The gauge group of the Left-Right symmetric models is
$SU(3)_C \times
U(2)_L\times SU(2)_R\times U(1)_{B-L}$.
In these models the gauge bosons from $SU(2)_L$ and $SU(2)_R$ will in general
mix. Due to this mixing at the one loop level CP violating
three gauge boson
form factors will be generated. The diagrams which contribute are shown in Fig.
3. For the lighter W and Z gauge bosons and the photon couplings to fermions
can be parametrized as
  $$ \eqalign {
L  =& -e\bar f_iQ_i\gamma_\mu f_iA^\mu - {e\over 2\sin\theta_W \cos\theta_W}
\bar f_i(g_V^{i}-  g_A^{i} \gamma_5)\gamma_\mu f_i Z^\mu \cr
+& \bar u_i
(f_V^{ij} - f_A^{ij}\gamma_5)\gamma_\mu d_j W^{+\mu} + H.C. \cr}\eqno(9) $$
Here $Q_i$ is the electric charge of fermions in units of $e >  0$, $g^i_V$ and
 $g^i_A$ are the coupling constants of $Z$ to neutral and axial neutral
 currents, and
 $$ \eqalign{
f_V^{ij} =& {e\over 2\sqrt{2}}({V_{Rij}\sin\zeta\over \sin\theta \cos\theta_W}
 -{V_{Lij}\cos\zeta\over \sin\theta_W})\;\cr
f_A^{ij} =& {e\over 2\sqrt{2}}(-{V_{Rij}\sin\zeta\over \sin\theta \cos\theta_W}
 - {V_{Lij}\cos\zeta\over \sin\theta_W}) \cr }\eqno(10) $$
Here $V_{R,L}$ are the right-handed and left-handed KM matrices, $\zeta$ is the
$W_L -W_R$ mixing angle, and $\sin\theta = g_{B-L}/\sqrt{g_R^2 + g_{B-L}^2}$
with
$g_i$ being the gauge couplings. In the limit $g_R = g_L$,
$\sin\theta \cos\theta_W = \sin\theta_W$.
At the one loop level non vanishing $f_4^Z$ and
$f_6^V$ will be generated while $f_7^V$ is still zero.
Evaluating the diagrams in Fig. 3, we obtain
  $$ \eqalign {
f_4^Z =& -{ N_C\over 16\pi^2 \cos^2\theta_W} 4m_{u_i}m_{d_j}
 {\rm Im}(f_V^{ij}f_A^{ij*})
(g_A^{d}I_{1d} - g_A^{u}I_{1u}) \cr
f_6^V =& {N_C\over 16\pi^2\cos^2\theta_W^2} 4m_{u_i}m_{d_j}
 {\rm Im}(f_V^{ij}f_A^{ij*})
(g_V^{u}I_{0u} + g_V^{d}I_{0d}) \cr }\eqno(11)$$
where $N_C$ is the number of color and,
  $$ \eqalign {
I_{1d} =& \int d[\alpha] {1-2\alpha_1\over -\alpha_1\alpha_2 q^2 -
(1-\alpha_3)\alpha_3 m_W^2
+(\alpha_1 +\alpha_2)m_d^2 + \alpha_3 m_u^2} \cr
I_{0d} =& \int d[\alpha] {1\over -\alpha_1\alpha_2 q^2
-\alpha_3(1-\alpha_3)m_W^2
+(\alpha_1 +\alpha_2)m_d^2 + \alpha_3m_u^2} \cr }\eqno(12) $$
Exchange $m_d$ and $m_u$ in (12) to obtain $I_{1u}$ and $I_{0u}$. From the
above expressions
we clearly see that the heavest internal fermions dominate the contribution.
If there are only three generations of quarks and leptons,
 the dominant contribution is from the third generation.
\par
Again if $q^2$ is big enough
an absorbtive part of the amplitude will be generated,
we have
  $$ \eqalign {
{\rm Im} f_4^Z =& {N_C\over 4\pi \cos^2\theta_W q^2\beta^2_W}m_tm_b
 {\rm Im} (f_V^{ij}f_A^{ij*})
\cr
 \times&(\beta_bg_A^d[-2+{A_b\over B_b} \ln{A_b +B_b\over A_b-B_b}]
 \theta (q^2 - 4m_b^2)
-\beta_tg_A^u[-2+{A_t\over B_t}
\ln{A_t +B_t\over A_t-B_t}]\theta(q_2-4m_t^2))\cr
{\rm Im} f_6^V =& -N_C{m_tm_b
 {\rm Im} (f_V^{ij}f_A^{ij*})\over 4\pi \cos^2\theta_Wq^2\beta_W}
\cr
 \times& (g_V^d\ln{A_b +B_b\over A_b - B_b} \theta(q^2 - 4m_b^2)
+ g_V^u \ln{A_t + B_t\over A_t -B_t} \theta(q^2 -4m_t^2)) \cr }\eqno(13) $$

where $A_b = m_b^2 +m_W^2 -m_t^2 -{1\over 2}q^2$, $B_b = {1\over 2}q^2 \beta_W
\beta_b$ with $\beta_i = \sqrt{1- {4m_i^2\over q^2}}$. $A_t$ and $B_t$ are
obtained by exchanging $m_b$ and $m_t$.
Changing $g_V^i$ to $2\cos ^2\theta_WQ^i$, one obtains the expression for
$f_6^\gamma$. Setting $q^2=0$
we find that our expression agree with that obtained in [7,9]. \par
We use $m_b=4.5$GeV and $m_t=130$GeV to obtain nummerical results for
the form factors in (11) and (13). \par
For $q^2=(200{\rm GeV})^2$:
   $$ \eqalign {
      f_4^Z & = N_C{\rm Im}(f_V^{tb}f_A^{tb*})\cdot ( -4.5-i2.4) \times
    10^{-5} \cr
      f_6^Z & = N_C{\rm Im}(f_V^{tb}f_A^{tb*})\cdot (2.0-i7.3) \times
     10^{-5} \cr
         f_6^\gamma &= N_C{\rm Im}(f_V^{tb}f_A^{tb*})\cdot  (1.04-i1.07)\times
     10^{-3} \cr } \eqno(14) $$ \par
For $q^2=(500{\rm GeV})^2$:
   $$ \eqalign {
      f_4^Z & = N_C{\rm Im}(f_V^{tb}f_A^{tb*})\cdot ( -4.3-i4.7) \times
    10^{-5} \cr
      f_6^Z & = N_C{\rm Im}(f_V^{tb}f_A^{tb*})\cdot (2.4+i5.2)\times
     10^{-5} \cr
     f_6^\gamma &= N_C{\rm Im}(f_V^{tb}f_A^{tb*})\cdot (2.5+i11.9) \times
     10^{-3} \cr } \eqno(15) $$
$\zeta$ is constrained to be less than $10^{-2}$[15]. We will
then have small values for
$f_i^V$ (less than $10^{-5}$), and it will be hard to see these form factors in
experiment.
However if a heavy fourth generation exists, these form factors can be
enhanced by a factor of $m_{4u}m_{4d}/m_tm_b$. Here $m_{4i}$ are the fourth
generation fermion masses.\par
The form factors discussed here can be measured at $e^+e^-$ colliders with $
 \sqrt{s}\ge  2m_W$ or at $p \bar p$ colliders. Such measurements are CP tests.
CP tests in the process $e^+e^- \rightarrow W^+W^-$ have been
discussed in [1,16,17,18]. At LEP the sensitivity for the
dispersive parts of the form factors can reach
to 0.1, and at the proposed NLC $10^{-2}$. However, a detailed analysis
with the absorptive parts is still lacking, and there may be some observables,
which can be optimized to get higher sensitivity as is done for the process
$e^+e^-\rightarrow t\bar t$[19]. We hope to return to this
point in the near future. \par\vskip 15pt
{\bf Acknowledgments}\par
This work is supported by Australia Research Council.
\par\vfil\eject
\centerline{\bf Reference}
\vskip 12pt \par

[1] K. Hagiwara, R. D. Peccei, D. Zeppenfeld and K. Hikasa, Nucl. Phys.
B282 \par \ \ \ \  (1987) 253 \par
 \ \ \ \   J. F. Gaemers and G. Gounaris, Z. Phys. C1 (1979) 259 \par
[2] K. Hagiwara, KEK Preprint 91--215, Plenary talk at the Workshop
on Physics \par
\ \ \ \ and Experiments with Linear Colliders, Saaviseld\"a,
Finland, 9--14 September 1991 \par
[3] W. J. Marciano and A. Queijeiro, Phys. Rev. D33 (1986) 3449\par
[4] A. De Rujula, M. B. Gavela, O. Pene and F. J. Vegas, Nucl. Phys.
 B357 \par \ \ \ \ (1991) 311 \par
[5] F. Boudjema, C. P. Burges, C. Hamzaoui and J. A. Robinson, Phys. Rev.
\par \ \ \ \  D43 (1991) 3683 \par
[6] C. P. Burges and D. London, Preprint McGill--92/05, UdeM--LPN--TH-84
\par
[7] D. Atwood, C. P. Burges, B. Iriwin and J. A. Robinson Phys. Rev. D42
 (1990) \par \ \ \ \ 3770 \par
[8] X. G. He and B. H. J. McKellar, Phys. Rev. D42 (1990) 3221\par
[9] D. Chang, W. Y. Keung and J. Liu, Nucl. Phys. B335 (1991) 295 \par
[10] G. C. Branco and M. N. Rebelo, Phys. Lett. B160 (1985) 117 \par
\ \ \ \ \ J. Liu and L. Wolfenstein, Nucl. Phys. B289 (1987) 1 \par
\ \ \ \ \ S. Weinberg, Phys. Rev. D42 (1990) 860 \par
[11] A. M\'endez and P. Pomarol, Phys. Lett. B272, (1991), 313 \par
[12] J. Gunion and D. Wyler, Phys. Lett. B248 (1990) 170 \par
\ \ \ \ \ D. Chang, W. Y. Keung and T. C. Yuan, ibid. B251 (1990) 608\par
[13] K. F. Smith et al., Phys. Lett. B234 (1990) 234\par
[14] X. G. He, B. H. J. McKellar and S. Pakvasa, Phys. Lett. B254 (1991)
 231 \par
[15] J. F. Donoghue and B. R. Holstein, Phys. Lett. B113 (1982) 382 \par
\ \ \ \ \ I. I. Bigi and J. M. Frere, Phys. Lett. B110 (1982) 255 \par
[16] G. Gounaris, D. Schildknecht and F. M. Renard, Phys. Lett. B263 (1991)
291 \par
[17] J. P. Ma Heidelberg--Preprint, HD--THEP-91-33 \par
[18] A. Bilal, E. Masso and A. De Rujula, Nucl. Phys. B355 (1991) 549\par
[19] D. Atwood and A. Soni, Phys. Rev. D45 (1992) 2405\par
\vfil
\eject
\par\noindent
Figure Caption: \par\noindent
Fig. 1. The Feynmann diagrams which contribute to $f_4^Z$ in the two-doublet
Higgs model.\par\noindent
Fig. 2a. The real part of the form factor $f_4^Z$ divided by $d_{11}
d_{12}d_{13}$ as function of $\Delta m=m_2-m_1$ in GeV with $m_C=60$GeV.
The solide line is for $q^2=(200{\rm GeV})^2$, the dashed line is for
$q^2=(500{\rm GeV})^2$. \par\noindent
Fig. 2b. The same as in Fig. 2a, but with $m_C=1$TeV. \par\noindent
Fig. 2c. The imaginary  part of the form factor $f_4^Z$ divided by $d_{11}
d_{12}d_{13}$ as function of $\Delta m=m_2-m_1$ in GeV with $m_C=60$GeV.
The solide line is for $q^2=(200{\rm GeV})^2$, the dashed line is for
$q^2=(500{\rm GeV})^2$.\par\noindent
Fig. 2d. The same as in Fig. 2c, but with $m_C=1$TeV. \par\noindent
Fig. 3. The Feynmann diagrams which contribute to
$f_4^Z$ and $f_6^V$ in the Left-Right symmetric model.\par
\vfil
\end